\documentclass[prd,aps,twocolumn,preprintnumbers,amsmath,amssymb,superscriptaddress]{revtex4}
\pdfoutput=1

\usepackage{amsmath}
\usepackage{graphicx}
\usepackage{dcolumn}
\usepackage{bm}
\usepackage{amssymb}
\usepackage{latexsym}
\usepackage{epstopdf}
\usepackage{threeparttable}
\usepackage{booktabs}
\usepackage{tabularx}
\usepackage{ulem}
\usepackage{mathrsfs}
\usepackage{hhline}
\usepackage{multirow}
\usepackage{color}
\usepackage[colorlinks,linkcolor=magenta,anchorcolor=blue,citecolor=green]{hyperref}

\def\be{\begin{equation}}
\def\ee{\end{equation}}
\def\bea{\begin{eqnarray}}
\def\eea{\end{eqnarray}}

\begin{document}
\title{Large $r$ against Trans-Planckian Censorship in Scalar-Tensor Theory?}

\author{Jiaming Shi}
\email{2016jimshi@mails.ccnu.edu.cn}
\affiliation{Institute of Astrophysics, Central China Normal University, Wuhan 430079, China}

\author{Taotao Qiu}
\email{qiutt@mail.ccnu.edu.cn}
\affiliation{Institute of Astrophysics, Central China Normal University, Wuhan 430079, China}
\affiliation{Key Laboratory of Particle Astrophysics, Institute of High Energy Physics, Chinese Academy of Sciences, Beijing 100086, China}
\begin{abstract}
In this paper, we discuss about the possibility to enhance the tensor-to-scalar ratio $r$ under the condition of Trans-Planckian censorship conjecture (TCC), thus $r\sim O(10^{-3})$ could be observable within the sensitivity of future experiments. We make use of the scalar-tensor theory where inflaton is nonminimally coupled to gravity. After demonstrating that the TCC condition could be modified in scalar-tensor theory, we show that due to the effects of modified gravity at the end of inflation, a large $r\sim O(10^{-3})$ could be allowed without violating the TCC. Moreover, the modification can give rise to a weak coupling of gravity to the inflation field. If such an effect has been present as early as inflation starts, it would imply that in our case, the Universe might have experienced an asymptotically safe period at its early time.
\end{abstract}

\maketitle
\section{introduction}
Inflationary cosmology \cite{Guth:1980zm,Albrecht:1982wi,Linde:1981mu,Linde:1983gd} can not only solve various problems in the Standard Big-Bang cosmology, but also provide a good explanation for the origin of structure in the universe. The key point is that the quantum fluctuations, initially generated inside the horizon, can grow with the universe's expansion and cross the Hubble horizon in the inflationary phase, and finally enter the horizon again for our observation. The superhorizon perturbations thus can leave hints on the Cosmic Microwave Background (CMB), which can be constrained by the observational data \cite{Akrami:2018odb}.

According to a general inflationary paradigm, at the very beginning, the fluctuations are constrained within a very small region.  If the duration of inflation is long enough and the region is below the Planck scale, say, $l_p=(M_p)^{-1}\approx 1.6\times 10^{-33}$cm, the quantum effects of those fluctuations will become robust. Since we don't have complete quantum theory of general relativity, it is difficult to make any predictions about those fluctuations, if they exit the horizon and make themselves observable to us. This is what has been called `` the Trans-Planckian Problem" proposed in \cite{Martin:2000xs}, see \cite{Brandenberger:2012aj} for a review.

Aiming at this problem, a ``Trans-Planckian Censorship Conjecture'' (TCC) has been proposed recently \cite{Bedroya:2019snp}, stating that the sub-Planckian fluctuations should never cross the Hubble horizon and remain quantum, therefore at least for the fluctuation modes that we observed today, the Trans-Planckian problem will never happen \cite{Bedroya:2019tba}. The conjecture, expressed in mathematical formulation, turns out to be
\be
\label{TCC}
e^N<\frac{M_P}{H_f}\,,
\ee
where $N$ is e-folding number of the inflation, $H_f$ is the Hubble parameter at the end of inflation, and $M_P$ is the reduced Planck mass. This can be regarded as a generalization of Penrose's cosmic censorship \cite{Penrose:1969pc}. As an implication to inflationary cosmology, investigations show that the TCC can set an upper bound on the inflationary Hubble scale $H_{\mathrm{inf}}$ and the tensor-to-scalar ratio $r$. For instance, ref. \cite{Bedroya:2019tba} has demonstrated that in the canonical single-field inflation model with a constant Hubble scale during inflation and the rapid reheating after inflation, the constraints are given by $H_{\mathrm{inf}}< 10^{-20}M_P$, and furtherly $r<10^{-31}$, via the observational value of the scalar power spectrum $P_\zeta\sim 10^{-9}$ \cite{Bedroya:2019tba}. The bound is tighter if the universe of the pre-inflationary period was dominated by radiation \cite{Brandenberger:2019eni}, whereas the upper bound on $r$ can be relaxed to $r<10^{-8}$ when the reheating process is considered \cite{Mizuno:2019bxy}. In $k$-inflation models when non-trivial sound speed is involved which can modulate the Hubble horizon, the bound on $r$ is also very tight $r<10^{-33}$ \cite{Lin:2019pmj}, while ref. \cite{Kadota:2019dol} shows that the upper bound could be affected by the choice of e-folding number $N$.

Such a stringent constraint on $r$, if satisfied, will make it difficult (if not impossible) to have it detected, as the sensitivity of the current and future experiments such as BICEP/Keck \cite{Ade:2018gkx}, AliCPT \cite{Li:2018rwc}, and LiteBIRD \cite{Matsumura:2016sri} and so on can only reach $10^{-3}$ at their furthest. On the other hand, if a large $r\sim O(10^{-3})$ were detected in the future experiments, does it imply that the TCC, at least its predictions on the inflationary cosmology, might not be correct, and we will have to face to the Trans-Planckian problem again?

There have been investigations on how to loose the constraints on $r$ by the TCC, so that the models with larger $r$ can also be compatible with this conjecture. For instance, the multistage inflation \cite{Mizuno:2019bxy,Torabian:2019qgl,Li:2019ipk,Torabian:2019zms} and constant roll inflation \cite{Kamali:2019gzr} can relax the bound to a limited degree, though the bound still does not reach the value within the sensitivity of the experiments. Another way to sufficiently improve $r$ under the TCC is to postulate that the quantum fluctuations originate from the non-Bunch-Davies (NBD) vacuum, which is not the lowest-energy state of de Sitter spacetime \cite{Brahma:2019unn}. In this paper, we will scrutinize this question again by considering inflationary models in scalar-tensor theory, where the inflaton field is nonminimally coupled to the Einstein gravity. Naively speaking, the nonminimal coupling will affect the spectrum by having the coupling function in its dominator, but will that also modify the TCC condition (\ref{TCC})? We will investigate whether the involvement of the nonminimal coupling will play a role in loosing the constraint on $r$ by the TCC.

The rest of the paper is organized as follows: in Sec. II we briefly review the constraint on $r$ by the TCC in the simplest case, which was initially obtained in \cite{Bedroya:2019tba}. In Sec. III we investigate such a constraint based on scalar-tensor theory, and in Sec. IV we come to our  discussions and conclusions.

\section{TCC, and its constraint on $H$ and $r$}
In this section, we will present how Eq. (\ref{TCC}) can give rise to the constraints on physical parameters such as $H$ and $r$ in inflation models. In \cite{Bedroya:2019tba}, two assumptions are taken: 1) that the Hubble parameter remains nearly constant during inflation, namely $H_i\simeq H_f$, where $H_i$ means the Hubble parameter at the beginning of the inflation; 2) that the reheating process will be instant and have no contribution on the number of e-folding. For current stage we will hold the assumption as well. For a fluctuation mode with the wavelength $a/k$ that exit the horizon at the initial time (denoted by the subscript ``$i$") and reenter into the horizon today (denoted by the subscript ``$0$"), where $k$ is its wave-number, we have $k=a_iH_i=a_0H_0$, and therefore
\be
\label{relation1}
\frac{H_i}{H_0}=\frac{a_0}{a_i}=\frac{a_f}{a_i}\frac{a_{rh}}{a_f}\frac{a_0}{a_{rh}}\,,
\ee
where the subscript ``$rh$'' denotes quantities at the end of reheating process. Note that $a_f/a_i=e^N$, and since we assume that the reheating process is negligible, we have $a_{rh}/a_f\simeq 1$. For $a_0/a_{rh}$, since we roughly have $a\sim T^{-1}$ where $T$ is the temperature of radiation, we have
\be
\frac{a_0}{a_{rh}}\simeq \frac{T_{rh}}{T_0}\simeq\frac{\rho^{1/4}_{rh}}{T_0}\simeq\frac{\rho^{1/4}_f}{T_0}\,,
\ee
where we consider $\rho_{rh}=g_\ast T_{rh}^4$, with the number $g_\ast$ neglected, and $\rho_{rh}\simeq\rho_{f}$, since reheating is fast and inflation has transferred totally to radiation. Also taking note that $\rho_f=3H_f^2M_P^2$, Eq. (\ref{relation1}) is written as
\be
\frac{H_i}{H_0}=\frac{1}{T_0}e^N(3H_f^2M_P^2)^{1/4}\,.
\ee
Taking into account the TCC inequality (\ref{TCC}) and considering $H_i\simeq H_f$ (denoted as $H$ thereafter), one has
\be
\label{Hconstraint}
\frac{H}{M_P}<3^{1/6}\left(\frac{H_0}{T_0}\right)^{2/3}\approx 10^{-20}\,,
\ee
where we take the values of the current Hubble parameter and CMB temperature, $H_0\approx 10^{-42}$GeV and $T_0\approx 10^{-13}$GeV, respectively. Moreover, by considering the slow-roll approximation $H\simeq \sqrt{V/3}/M_P$ where $V$ is the potential of inflaton, we can further constraint $V$ to be $V<10^{-38}M_P^4$.

On the other hand, inflation generates primordial perturbations which can predict scale-invariant scalar power spectrum $P_\zeta$, as well as tensor spectrum, featured by its ratio to the scalar one, namely the scalar ratio $r$. For k-essence inflation models without any higher derivatives or nonminimal coupling, $P_\zeta$ and $r$ can be expressed in a general form:
\be
\label{Pandr}
P_\zeta=\frac{H^2}{8\pi^2M_P^2\epsilon c_s}\,,~~~r=16\epsilon c_s\,.
\ee
The current observation requires $P_\zeta\sim 10^{-9}$, and combining with the constraint on $H$ given above, one can have the constraint on $r$, which is
\be
\label{rconstraint}
r<\frac{2}{\pi^2}\times 10^{-31}\,.
\ee

This gives a very stringent constraint on inflation models. However, it was stated in \cite{Mizuno:2019bxy} that the constraint could be loosen when one of the two conditions is abandoned. In \cite{Mizuno:2019bxy}, the authors considered that the reheating process is no longer instantaneous, therefore the relation $a_{rh}\simeq a_f$ and $\rho_{rh}\simeq \rho_f$ are no longer applicable. Moreover, assuming $\rho\sim a^{-3(1+w)}$ during the reheating time, we have
\be
e^{N_{rh}}=\frac{a_{rh}}{a_f}=\left(\frac{\rho_f}{\rho_{rh}}\right)^\frac{1}{3(1+w)}=\left(\frac{\sqrt{3}M_PH_f}{T_{rh}^2}\right)^\frac{2}{3(1+w)}\,.
\ee
Inserting this into Eq. (\ref{relation1}) we have:
\be
\frac{H}{M_P}<\frac{M_PH_0}{T_{rh}T_0}\left(\frac{M_PH}{T_{rh}}\right)^{\alpha-1}\,,~\alpha\equiv\frac{2}{3(1+w)}\,.
\ee
For $w=-1/3$ ($\alpha=1$) and $T_{rh}\simeq 1\text{MeV}$, as the authors in \cite{Mizuno:2019bxy} have chosen, $H<10^{-8}M_P$ can be obtained. Moreover, from the definition of $r$ in Eq. (\ref{Pandr}), one can get the constraint on $r$:
\be
\label{rconstraint2}
r<10^{-8}\left(\frac{1\text{MeV}}{T_{rh}}\right)^2\,.
\ee
One can see that, with the addition of reheating process, the constraint on both $H$ and $r$ has been loosen very much, but still far from the bounds that could be reached by the current experiments. On the other hand, although the condition of instant reheating has been abandoned, one still requires $T_{rh}\gtrsim 1$MeV, as has been required by the big bang nucleosynthesis (BBN) constraints.

\section{TCC in Scalar-tensor Theory}
\label{sec3}
\subsection{Inflation models in scalar-tensor theory}
In this subsection, we consider the inflation models in scalar-tensor theory. The action of scalar-tensor theory is given by
\begin{align}
\label{action}
S=\int  d^{4}x\sqrt{-g}\left[\frac{M_P^{2}}{2}f(\phi)R+P(\phi,X)\right]\,,
\end{align}
where $R$ is the Ricci scalar, $X=-\nabla_\mu\phi\nabla^\mu\phi/2$ is the kinetic term, $f(\phi)$ is an arbitrary function of $\phi$ that couples to the Einstein gravity. The coupling can also be viewed as the varying of the Planck mass or gravitational coupling constant, via $G(t)=G_N/f(\phi)$ with $G_N$ being the Newtonian gravitational constant \cite{Damour:1990tw}. When $G\sim\phi^{-1}$, it is reduced to the famous Brans-Dicke theory \cite{Brans:1961sx}, which is the first example of the scalar-tensor theory.

We consider the FRW metric
\begin{align}
\label{metric}
ds^2=-dt^2+a(t)^2\delta_{ij}dx^idx^j\,,
\end{align}
with $a(t)$ the scale factor and the Hubble parameter $H=\dot{a}/a$, the Friedmann equations are
\bea
\label{eom1}
3fM_{P}^{2}H^{2}&=&2XP_{X}-P-3M_{P}^{2}\dot{f}H\,,\\
\label{eom2}
-2fM_{P}^{2}\dot{H}&=&2XP_{X}+M_{P}^{2}\ddot{f}-M_{P}^{2}\dot{f}H\,,
\eea
and the equation of motion of scalar field is
\begin{align}\label{eom3}
&(2XP_{XX}+P_{X})\ddot{\phi}+3P_{X}H\dot{\phi}+2P_{X\phi}X\\&
-P_{\phi}-3M_{P}^{2}(\dot{H}+2H^{2})f_{\phi}=0\,.
\end{align}
Here we have used derivative form $f_x\equiv\partial f/\partial x$.

In order to discuss about the perturbations, we first perturb the metric (\ref{metric}) in the ADM form
\cite{{DeFelice:2011uc,Kobayashi:2011nu}}:
\begin{align}
\label{ADM}
ds^2=-N^2dt^2+\gamma_{ij}(dx^i+N^idt)(dx^j+N^jdt)
\end{align}
with the three dimensional metric
\begin{align}
\label{gij}
\gamma_{ij}=a^{2}e^{2\zeta}(\delta_{ij}+h_{ij}+\frac{1}{2}h_{ik}h_{kj})
\end{align}
and
\begin{align}
\label{Ni}
N=1+\alpha\,,N_{i}=N^j\gamma_{ij}=\partial_{i}\beta\,.
\end{align}
Here $\alpha,\beta,\zeta$ are scalar perturbations and $h_{ij}$ is a tensor perturbation satisfying $h_{ii}=\partial_jh_{ij}=0$. We choose the uniform field gauge $\delta\phi=0$, thus the second order action for the scalar perturbations contains variables $s=a,\alpha,\beta,\zeta$ and their higher order derivatives, namely,
\begin{align}
\label{S2}
S_{S}^{(2)}=\intop d^{4}x\mathcal{L}(s,\partial s,\partial^{2}s,\partial^{3}s)\,.
\end{align}

Varying the second order action \eqref{S2} with respect to $\alpha$ and $\beta$, two corresponding equations of motion can be obtained. Finally the solutions to $\alpha$ and $\beta$ are
\begin{align}
\label{ab}
 \alpha=\frac{\dot{\zeta}}{H+\frac{\dot{f}}{2f}}\,,~\beta=-\frac{\zeta}{H+\frac{\dot{f}}{2f}}+\chi\,,~\partial^{2}\chi=\frac{a^{2}\dot{\zeta}\epsilon_{s}}{c_{s}^{2}}\,,
\end{align}
with the parameter
\begin{align}
\label{e}
\epsilon_{s}=\frac{fXP_{X}+3M_{P}^{2}\dot{f}^{2}/4}{\left(fH+\dot{f}/2\right)^{2}M_{P}^{2}}\,,
\end{align}
and the sound speed of scalar modes
\begin{align}
\label{cs2}
c_{s}^{2}=\frac{fXP_{X}+3M_{P}^{2}\dot{f}^{2}/4}{2fX^{2}P_{XX}+fXP_{X}+3M_{P}^{2}\dot{f}^{2}/4}\,.
\end{align}
Eliminating $\alpha$ and $\beta$ in the action \eqref{S2}  with the help of Eq.\eqref{ab}, we obtain the quadratic action only depending on the curvature perturbation $\zeta$
\begin{align}
\label{S20}
S_{S}^{(2)}=\intop d^{4}xa^{3}\left[\mathcal{G}_{S}\dot{\zeta}^{2}-\frac{\mathcal{F}_{S}}{a^{2}}(\partial\zeta)^{2}\right]\,,
\end{align}
where
\begin{align}
\label{GS}
\mathcal{G}_{S}=M_{P}^{2}f\epsilon_{s}/c_{s}^{2}\,,
\mathcal{F}_{S}=M_{P}^{2}f\epsilon_{s}\,.
\end{align}
Define variable $u=z\zeta$ with $z=a\sqrt{2\mathcal{G}_{S}}$, the Mukhanov-Sasaki equation in the Fourier space can be obtained from (\ref{S20}) as:
\begin{align}
\label{MSeq}
u_k''+c_{s}^{2}k^{2}u_k-\frac{z''}{z}u_k=0\,,
\end{align}
where a prime denotes differentiation with respect to conformal time $\eta=\int dt/a$ and $k$ is a comoving wave number. Eq.\eqref{MSeq} can be solved to get the curvature perturbation $\zeta$, and its power spectrum at the Hubble horizon crossing is therefore given by
\begin{align}
\label{ps}
\mathcal{P}_{S}(k)\equiv\frac{k^{3}}{2\pi^{2}}|\zeta_{k}|^{2}
=\frac{H^{2}}{8\pi^{2}\mathcal{G}_{S}c_{s}^{3}}
=\frac{H^{2}}{8\pi^{2}M_{P}^{2}f\epsilon_{s}c_{s}}\,.
\end{align}

Similarly, the quadratic action for the tensor perturbations is given by
\begin{align}
\label{ST}
S_{T}^{(2)}=\frac{1}{8}\intop d^{4}xa^{3}\left[\mathcal{G}_{T}\dot{h}_{ij}^{2}-\frac{\mathcal{F}_{T}}{a^{2}}(\partial h_{ij})^{2}\right]\,,
\end{align}
where
\begin{align}
\label{GT}
\mathcal{G}_{T}=M_{P}^{2}f\,,
\mathcal{F}_{T}=M_{P}^{2}f\,.
\end{align}
The power spectrum of the primordial tensor perturbation $h_{ij}$ is
\begin{align}
\label{pt}
\mathcal{P}_{T}(k)=2\frac{k^{3}}{2\pi^{2}}|h_{k}|^{2}
=\frac{2H^{2}}{\pi^{2}\mathcal{G}_{T}c_{T}^{3}}
=\frac{2H^{2}}{\pi^{2}M_{p}^{2}fc_{T}}\,,
 \end{align}
where in our case,  the speed of primordial gravitational wave is $c_T^2=1$. Utilizing Eqs.\eqref{ps} and \eqref{pt}, the tensor-to-scalar ratio $r$ turns out to be
\begin{align}
\label{r}
r=\frac{\mathcal{P}_{T}(k)}{\mathcal{P}_{S}(k)}=16\epsilon_{s}c_{s}\,.
\end{align}

In fact, the action (\ref{action}) can also be rewritten in a GR-like form:
\be
S=\int  d^{4}x\sqrt{-g}\left[\frac{{\cal M}_P^2}{2}R+P(\phi,X)\right]\,,
\ee
with the definition: ${\cal M}_P\equiv M_P\sqrt{f}$. Here ${\cal M}_P$ is the effective Planck mass which also gets the nonminimal coupling term $f$ involved. Under this definition, the expression of scalar and tensor power spectra can also be rewritten as:
\be
\mathcal{P}_{S}(k)=\frac{H^{2}}{8\pi^{2}{\cal M}_P^2\epsilon_{s}c_{s}}~,~~~\mathcal{P}_{T}(k)=\frac{2H^{2}}{\pi^{2}{\cal M}_P^2c_{T}}\,,
\ee
while the tensor-to-scalar ratio $r$ remains unchanged.

\subsection{Modified TCC and its constraint on $r$}
In the scalar-tensor theory, we first argue that, the condition of the TCC in Eq. (\ref{TCC}) will be modified as:
\be
\label{TCC2}
e^N<\frac{{\cal M}_{Pi}}{H_f}=\frac{M_P\sqrt{f_i}}{H_f}\,,
\ee
where $f_i$ denotes the value of $f$ at the initial time. Note that, since the effective Planck mass is now a varying function rather than a constant, its value at different time would no longer be the same, therefore it is necessary to clarify that at which time the value is taken. Here we conclude that the value of ${\cal M}_{P}$ appearing here should be taken at the initial time of inflation. To see this, note that the model in (\ref{action}) can be cast into a minimal coupling model in its so-called Einstein frame, under such a conformal transformation:
\be
\hat{a}=\sqrt{f}a~,\hat{H}=\frac{H}{\sqrt{f}}(1+\frac{\dot f}{2Hf})\,,
\ee
where $\hat{}$ denotes quantities in Einstein frame, and when $f$ does not vary much, $\dot f/(2Hf)$ can be ignored so as we approximately have $\hat{H}\simeq H/\sqrt{f}$. In Einstein frame, the usual TCC (\ref{TCC}) is obtained by the fact that the fluctuation mode $k_{\mathrm{TP}}$ with the wavelength at the initial time ($\hat{a}_i/k_{\mathrm{TP}}=\hat{l}_{Pi}$) does not exceed the horizon at the final time: $\hat{a}_f/k_{\mathrm{TP}}=(\hat{a}_f/\hat{a}_i)\hat{l}_{Pi}<1/\hat{H}_f$. Here $\hat{l}_P\equiv 1/M_P$ is the Planck length, and $\hat{l}_{Pi}=\hat{l}_{Pf}=\hat{l}_{P}$ since in Einstein frame, Planck length and Planck mass are both constant. Transforming this formulation into Jordan frame, one gets
\be
\frac{a_f}{k_{\mathrm{TP}}}=\frac{a_f}{a_i}l_{Pi}=\frac{\hat{a}_f}{\hat{a}_i}\frac{\hat{l}_{Pi}}{\sqrt{f_f}}<\frac{1}{\hat{H}_f\sqrt{f_f}}=\frac{1}{H_f}\,,
\ee
where $f_f$ denotes the value of $f$ at the end of inflation, and since the Planck length in Jordan frame $l_{Pi}=1/{\cal M}_{Pi}$, Eq. (\ref{TCC2}) is obtained. Therefore we for the first time claimed that it is actually the initial value of Planck mass that matters in the TCC condition.  In the modified TCC Eq. (\ref{TCC2}), the effective Planck mass $\mathcal{M}_P$ can be regarded as the UV cut-off.

According to the new TCC condition (\ref{TCC2}) as well as the analysis in previous section, one can get the constraint on $H$ to be:
\be
\label{Hconstraint2}
\frac{H}{M_P\sqrt{f_i}}<\frac{H_0T_0}{H_fT_{rh}}e^{N_{rh}}\,,
\ee
where $N_{rh}=\ln(a_{rh}/a_f)$.

It is useful to change $N_{rh}$ to a more practical quantities such as $\rho$, $H$ and $T$, however, some subtleties are in order. In minimal coupling case where $f=1$, $a$ and $\rho$ are related by the energy conservation equation, which leads to $\rho\sim a^{-3(1+w)}$. Moreover, we have $\rho_f=3H^2M_P^2$, $\rho_{rh}={\cal O}(1)\times T_{rh}^4$. In our case, however, due to the presence of function $f$ in the Friedmann equations (\ref{eom1}) and (\ref{eom2}), the energy conservation equation will in general be modified, but if we assume that during reheating time $f$ does not vary much, $\rho\sim a^{-3(1+w)}$ can still hold. Furthermore, at the end of inflation we have $\rho_f=3H^2M_P^2f_f$ instead. $\rho_{rh}={\cal O}(1)\times T_{rh}^4$ still holds, but here $T_{rh}$ is related to its Einstein frame counterpart by the relation $T_{rh}=\sqrt{f_{rh}}\hat{T}_{rh}\simeq\sqrt{f_f}\hat{T}_{rh}$. Therefore Eq. (\ref{Hconstraint2}) becomes:
\be
\label{Hconstraint3}
\frac{H}{M_P\sqrt{f_i}}<\frac{H_0M_P\sqrt{f_f}}{T_0T_{rh}}\,,
\ee
for $w=-1/3$.

For $f_f=1$, the above constraints will reduce to the same results shown in \cite{Mizuno:2019bxy}. Although there is a difference of $f_i$, considering the compensating $f_i$ in power spectrum (\ref{ps}), there will be no effects in improving the constraints on $r$. However, in the presence of nontrivial $f_f$, things will be different as one more degree of freedom gets involved. In our case, the constraint on $r$ turns out to be:
\be
r<10^{-8}f_f\left(\frac{1\text{MeV}}{T_{rh}}\right)^2\,.
\ee
This is our main result. This demonstrates that, by admitting a non-trivial modification of gravity, especially at the end of inflation, it is possible to furtherly reduce $r$. For instance, the $r\gtrsim 10^{-3}$ could be allowed as long as we set
\be
f_f\gtrsim 10^{5}\,.
\ee

\begin{figure}[htb]
\centering
\includegraphics[scale=0.35]{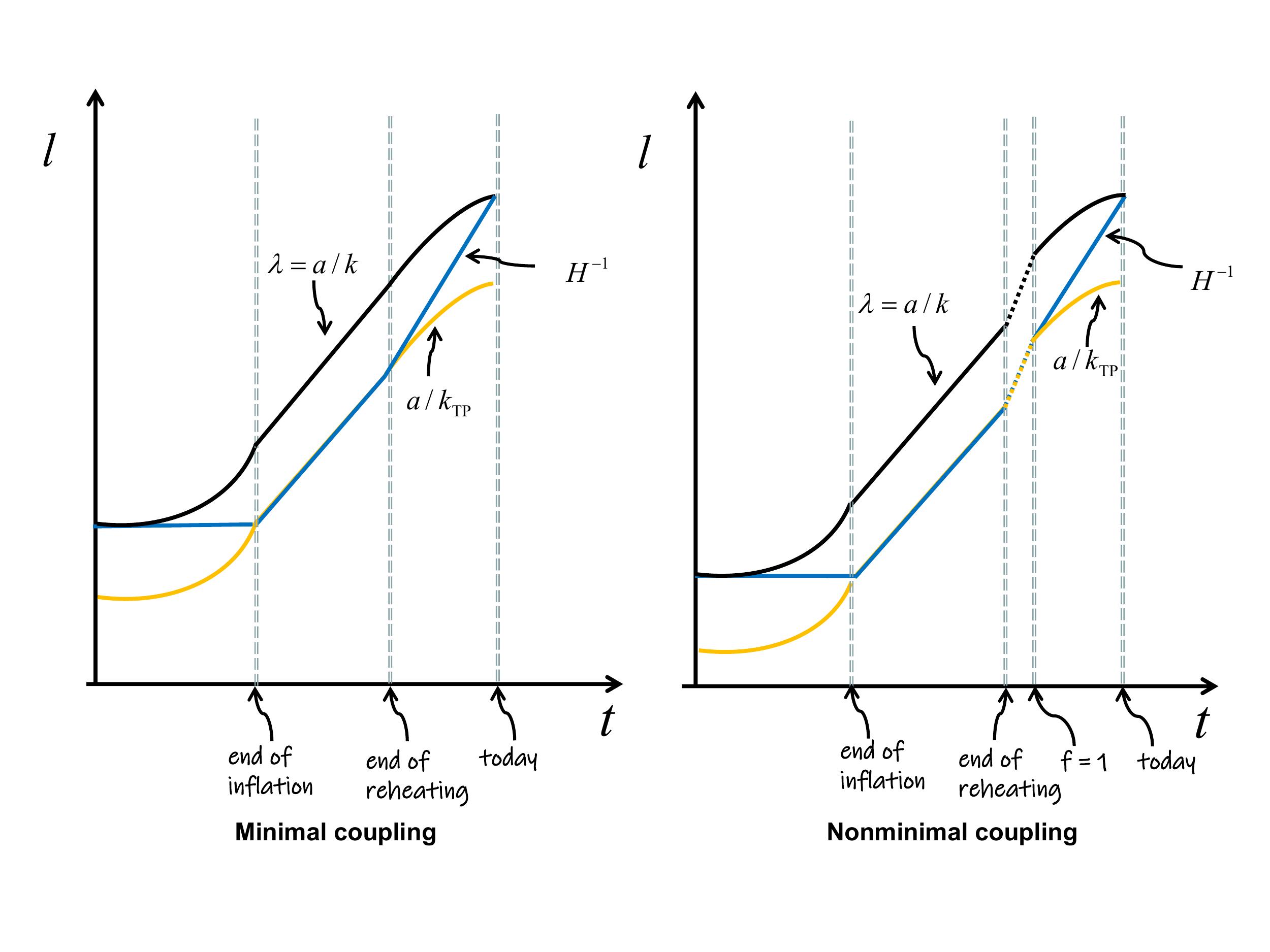}\\
\caption{The evolution of Hubble horizon, the fluctuation mode observed today and the fluctuation mode corresponding to Planck length are plotted in blue, black and yellow lines, respectively. The left and right panel corresponds to the GR case (minimal coupling to gravity) and Scalar-Tensor theory (nonminimal coupling to gravity). From the plot we can see that, in the early time where nonminimal coupling function $f$ is large, the three lines in the right one are all lowered compared to the left one. After reheating, it is assumed that there is a period where $f$ reduces to 1, recovering to the GR case. }
\label{sketch}
\end{figure}

To understand this scenario more clearly, we plot the evolution of fluctuation modes as well as the Hubble horizon in Fig. \ref{sketch}. We can see that, comparing to the case of GR, both lines of $H^{-1}$ and $l_P$ has been lowered in scalar-tensor theory, due to the correction of function $f$. Therefore the Hubble parameter satisfying TCC will get enlarged. This situation will last till the reheating process is completed. However, after the reheating, in order to have the Universe return to GR case, there need to be a period where the function $f\rightarrow 1$. It is also interesting to discuss about the mechanism that reduces $f$, although it is beyond the scope of the current discussion and we will postpone it to a future work.

Our mechanism can also be interpreted in its Einstein frame, where the universe behaves like GR. In this frame, the effect of our mechanism is equivalent to an even lower ``effective" reheating temperature $\hat{T}_{rh}=T_{rh}/\sqrt{f}$. As has been shown in (\ref{rconstraint2}), a lower reheating temperature is useful for improving the constraints on $r$. However, since the real reheating temperature is that in the Jordan frame, which is set to be larger than 1MeV, there is no need to worry about the conflict with BBN results.

As a side remark, note that the effective gravitational coupling constant in the scalar-tensor theory is defined as $8\pi G\sim 1/{\cal M}_P^2=1/(M_P^2f)$. Therefore $G$ is now running as $f$ varies with time. Moreover, for $f_f\gg 1$ , we will get $G_f\ll 1$, which means that the gravitational coupling will be very weak, at least at the end of inflation. If the weak coupling continues up to the beginning of the inflation, namely $f_i\simeq f_f$, this means, the universe might have experienced an asymptotically safe behavior in the early time. In this case, $G$ is running with respect to energy scale due to Renormalization Group Equations \cite{Weinberg:2009bg, Weinberg:2009wa, Cai:2012qi, Cai:2013caa}, while in late time when $f\rightarrow 1$ the usual result of GR is recovered. It gives the hint that, a gravity theory with asymptotically safe behavior at the early time might help to reconcile the TCC conjecture with the detectability of tensor to scalar ratio from the observations. On the other hand, if it is verified to be so, it will be another proof that the theory of asymptotic safety can work as a good candidate of UV-complete theory.

\section{Conclusion and discussion}
\label{sec4}
Recently it has been pointed out that in order to have robust quantum effects of fluctuation not affect standard cosmological scenario, a condition of Trans-Planckian Censorship Conjecture has to be satisfied, namely sub-Planckian quantum fluctuations should remain quantum. According to the condition, it has been found that in usual case, the tensor-to-scalar ratio has been severely constrained, which couldn't reach the regions of detectability of the current observations.

In this paper, we resort to a scalar-tensor theory for improving the value of the tensor-scalar ratio in the TCC. First of all, we showed that, within the framework of scalar-tensor theory, the expression for the TCC might have to be modified. This is because the effective Planck mass in scalar-tensor theory is no longer $M_P$ but ${\cal M}_P=M_P\sqrt{f}$, and what plays the role is the value of ${\cal M}_P$ in its initial value. Then, we rederive the TCC condition. We find that, the constraint on Hubble parameter by the TCC will be modified by two parameters, $f_i$ and $f_f$. Although the first one could be compensated by the $f_i$ presented in the expression of power spectrum and thus has no real effects on constraining $r$, the latter does make sense. By properly choosing the value of $f_f$, it is possible that the constraint on $r$ gets relaxed.

In our opinion, such an improvement may be because that, for a large value of $f$, both Hubble horizon and the wavelength starting from Planck length will get lowered, then the upper bound of $H_i$ (modulated by $\sqrt{f_i}$ as well) will get raised. If this scenario is to be true, some mechanism after the reheating has completed may be needed for the $f$ function to be reduced to unity, so that our Universe can return to GR. In Einstein frame, however, our mechanism is equivalent to having an even lower reheating temperature.

It is also important to investigate the observational constraints on the large $f_f$. For example, \cite{Alvey:2019ctk} has provided the constraints on $f$ from the big bang nucleosynthesis (BBN) observations, implying that $f_{\mathrm{BBN}}\simeq1$. However, the constraint on $f$ for even earlier time is still quite unclear. Once we get the constraint on $f$ at the early time of universe such as reheating time, we should be able to testify our mechanism in a more practical way.

Moreover, the largeness of $f$ can also lead to a smallness of the gravitational coupling constant $G$, namely a weak coupling of gravity to the inflaton field. If such a weak coupling remains backwards in time, it implies that we may have experienced an asymptotically safe behavior at early time. As a prosperous candidate of UV-complete theory, it is also reasonable to expect that an asymptotically safe universe would provide a solution of TCC-related problems. A concrete model of realizing it will be discussed in the forthcoming work.

As a final remark, let's mention that it is also argued in some work (e.g.\cite{Berera:2020dvn} ) that the TCC could be refined with some $O(1)$ factor induced by the swampland distance conjecture (SDC), which is also useful to relax the constraints on $r$. However, as they concluded, such a refinement is of limited significance. On the other hand, our approach had provided a possibility that the TCC can be greatly refined with the factor $f$. Moreover, it is also interesting to consider the combination of the two effects, namely the SDC effect on scalar-tensor theory. These topics are also postponed to the future \footnote{We thank the anonymous referee to point out this interesting work to us.}.

$\newline$
\begin{acknowledgments}
We thank Zheng Fang, Yun-Song Piao, Yi-Fu Cai and Mingzhe Li for helpful discussions. This work was supported by the National Natural Science Foundation of China under Grants No.~11653002 and No.~11875141.
\end{acknowledgments}

\bibliographystyle{apsrev4-1}
\bibliography{TCCGW2020}
\end{document}